\documentclass[secnumarabic,onecolumn,noshowpacs,noshowkeys,10pt]{revtex4}
\usepackage{amsfonts}
\usepackage{amsmath}
\usepackage{graphicx}
\usepackage{psfig}
\usepackage{dcolumn}
\usepackage{bm}

\def\lessim{\lower.5ex\hbox{$\; \buildrel < \over \sim \;$}}
\def\gtrsim{\lower.5ex\hbox{$\; \buildrel > \over \sim \;$}}
\input{tcilatex}
\begin{document}

\preprint{}
\title{Finite-Size Effects and Scaling for the Thermal QCD Deconfinement
Phase Transition within the Exact Color-Singlet Partition Function}
\author{M. Ladrem and A. Ait-El-Djoudi}
\affiliation{Laboratoire de Physique des Particules et Physique Statistique\\
Ecole Normale Sup\'{e}rieure-Kouba,\\
B.P. 92, 16050, Vieux-Kouba, Algiers, Algeria.\\
E-mail: Ladrem@eepad.dz, Aitamel@eepad.dz }
\date{\today }

\begin{abstract}
\textbf{Abstract.}

We study the finite-size effects for the thermal QCD Deconfinement Phase
Transition (DPT), and use a numerical finite size scaling analysis to
extract the scaling exponents characterizing its scaling behavior when
approaching the thermodynamic limit $\left( V\longrightarrow \infty \right) $%
. For this, we use a simple model of coexistence of hadronic gas and
color-singlet Quark Gluon Plasma (QGP) phases in a finite volume. The
Color-Singlet Partition Function (CSPF) of the QGP cannot be exactly
calculated and is usually derived within the saddle point approximation.
When we try to do calculations with such an approximate CSPF, a problem
arises in the limit of small temperatures and/or volumes $\left(
VT^{3}<<1\right) $, requiring then additional approximations if we want to
carry out calculations. We propose in this work a new method for an accurate
calculation of any quantity of the finite system, without explicitly
calculating the CSPF itself and without any approximation. By probing the
behavior of some useful thermodynamic response functions on the hole range
of temperature, it turns out that in a finite size system, all singularities
in the thermodynamic limit are smeared out and the transition point is
shifted away. A numerical finite size scaling analysis of the obtained data
allows us to determine the scaling exponents of the QCD DPT. Our results
expressing the equality between their values and the space dimensionality is
a consequence of the singularity characterizing a first order phase
transition and agree very well with the predictions of other FSS theoretical
approaches and with the results of both\ lattice QCD and Monte Carlo models
calculations.
\end{abstract}

\maketitle

\vspace{0.5cm}

%\section{Introduction}
%%%%%%%%%%%%%%%%%%%%%%%%

\section{Introduction}

It is generally believed that at sufficiently high temperatures and/or
densities a new phase of matter called the Quark Gluon Plasma (QGP) can be
created. This is logically a consequence of the quark-parton level of the
matter structure and of the dynamics of strong interactions described by the
Quantum Chromodynamics (QCD) theory. Its existence, however, has been partly
supported by lattice QCD calculations and the Cosmological Standard Model
predictions. The only available way it seems to study experimentally the QCD
Deconfinement Phase Transition (DPT) is to try to create in
Ultra-Relativistic Heavy Ion Collisions (URHIC) conditions similar to those
in the early moments of the Universe, right after the Big Bang. If ever the
QGP is created in URHIC, the volume within which the eventual formation
would take place would certainly be finite. Also, in lattice QCD studies,
the scale of\ the lattice space volume is finite. This motivates the study
of the Finite Size Effects (FSE) on the expected DPT from a Hadronic Gas
(HG) phase to a QGP phase. These effects are certainly important since
statistical fluctuations in a finite volume may hinder a sharp transition
between the two phases. Phase transitions are known to be infinitely sharp,
signaled by some singularities, only in the thermodynamic limit \cite%
{LeeYang52}. In general, FSE lead to a rounding of these singularities as
pointed out in \cite{Im80,BL84}. However, even in a such situation, it is
possible to obtain informations on the critical behavior. Large but finite
systems show a \textit{universal} behavior called \textquotedblleft
Finite-Size Scaling\textquotedblright\ (FSS), allowing the put of all the
physical systems undergoing a phase transition in a certain number of
universality classes. The systems in a given universality class display the
same critical behavior, meaning that certain dimensionless quantities have
the same values for all these systems. \textit{Critical exponents} are an
example of these universal quantities.

In the present work, we study the finite size behavior for the thermally
driven QCD DPT. For this purpose, we consider a simple Phase Coexistence
Model (PCM) used in \cite{SSG98}, in which the mixed phase system has a
finite volume: $V=V_{HG}+V_{QGP}$. The parameter $\mathfrak{h}$ representing
the fraction of volume occupied by the HG: $V_{HG}=\mathfrak{h}V,$\ is then
defined and can be considered as an order parameter for the QCD DPT.
Assuming non-interacting phases (separability of the energy spectra of the
two phases), and using the PCM, the total partition function of the system
can be written as:%
\begin{equation}
Z(\mathfrak{h})=Z_{QGP}(\mathfrak{h})Z_{HG}(\mathfrak{h})Z_{Vac}(\mathfrak{h}%
),
\end{equation}%
where:%
\begin{equation}
Z_{Vac}(T,\mathfrak{h})=\exp (-BV_{QGP}/T)=\exp (-BV(1-\mathfrak{h})/T),
\end{equation}%
accounts for the confinement of the color charge by the real vacuum pressure
exerted on the perturbative vacuum $\left( B\right) $ of the bag models.

The mean value of any thermodynamic quantity of the system $\langle A(T,\mu
,V)\rangle $, as defined in \cite{SSG98}, can then be calculated by:%
\begin{equation}
\langle \mathcal{A}(T,\mu ,V)\rangle =\frac{\int\limits_{0}^{1}\mathcal{A}%
\left( \mathfrak{h},T,\mu ,V\right) Z\left( \mathfrak{h}\right) d\mathfrak{h}%
}{\int\limits_{0}^{1}Z\left( \mathfrak{h}\right) d\mathfrak{h}},
\end{equation}%
where $\mathcal{A}(\mathfrak{h},T,\mu ,V)$ is the total thermodynamic
quantity in the state $\mathfrak{h}$, given in the case of an extensive
quantity by: 
\begin{equation}
\mathcal{A}(\mathfrak{h},T,\mu ,V)=\mathcal{A}_{HG}(T,\mu ,\mathfrak{h}V)+%
\mathcal{A}_{QGP}(T,\mu ,(1-\mathfrak{h})V),
\end{equation}%
and in the case of an intensive quantity, by: 
\begin{equation}
\mathcal{A}(\mathfrak{h},T,\mu ,V)=\mathfrak{h}\mathcal{A}_{HG}(T,\mu ,%
\mathfrak{h}V)+(1-\mathfrak{h)}\mathcal{A}_{QGP}(T,\mu ,(1-\mathfrak{h})V),
\end{equation}%
with $\mathcal{A}_{QGP}$ and $\mathcal{A}_{HG}$ the contributions relative
to the individual QGP and HG phases, respectively.

We think that the PCM used in our present work is not a characteristic
signature in a finite volume which anticipates the behavior of a first order
phase transition in the TL. It is well known that finite size effects lead
to a non obvious order of the phase transition. A characteristic feature of
a truly first order transition in the TL can then appear in the second order
transition when the volume of the system is finite and vice versa, and this
is considered as a finite size artifact. For example, the phenomenon of
coexistence of phases can appear in a truly second order phase transition
when the volume of the system is finite, and we can also have a divergent
correlation length in a truly first order phase transition \cite{Or96}.
Certainly, these non-characteristic features must disappear when approaching
the TL.

For the HG phase, we have considered just pionic degrees of freedom, and the
partition function for such a system is simply given by:%
\begin{equation}
Z_{HG}=e^{\frac{\pi ^{2}}{30}T^{3}V_{HG}}.
\end{equation}

For the QGP phase, we have considered a free gas of quarks and gluons with
the exact color-singletness requirement. The color-singlet partition
function of the QGP derived using the group theoretical projection technique
formulated by Turko and Redlich \cite{RT81} can not be exactly calculated
and is usually calculated within the saddle point approximation in the limit 
$V_{QGP}T^{3}>>1$ as in \cite{EGR, TLR, MSS}. It turns out that the use of
the obtained approximated partition function for the calculation of a mean
value within the definition (3), has as a consequence the absence of the DPT 
\cite{YG02}. This is due to the fact that the approximation used for the
calculation of the color-singlet partition function breaks down at $%
V_{QGP}T^{3}<<1$, and this limit is attained in our case. This has been
emphasized in further works in which additional approximations have been
used to carry out calculations, as in \cite{TLR,SSG98}. We propose in the
following a new method which allows us to accurately calculate physical
quantities describing well the QCD DPT at finite volumes, avoiding then the
problem arising at $V_{QGP}T^{3}<<1$ and without any approximation. We
proceed by using the exact definition of the color-singlet partition
function for $Z_{QGP}$, without explicitly calculating it, in the definition
(3) of the mean value of a physical quantity. A first analytical step in the
calculation of the mean value is then achieved, and an expression with
integral coefficients is obtained. The double integrals are then carried out
with a suitable numerical method at each value of temperature and volume,
and the behavior of the physical quantity of the finite system can so be
obtained on the hole range of temperature, for various volumes, without any
restriction. Afterwards, scaling critical exponents characterizing the
scaling behavior of some quantities are determined using a \ numerical FSS
analysis.

\section{Exact Color-singlet partition function of the QGP}

The exact partition function for a color-singlet QGP contained in a volume $%
V_{QGP},$ at temperature $T$ and quark chemical potential $\mu ,$ is
determined by \cite{EGR}:%
\begin{equation}
Z_{QGP}(T,V_{QGP},\mu )=\frac{8}{3\pi ^{2}}\int\limits_{-\pi }^{+\pi
}\int\limits_{-\pi }^{+\pi }d\left( \tfrac{\varphi }{2}\right) d\left( 
\tfrac{\psi }{3}\right) M(\varphi ,\psi )\widetilde{Z}(T,V_{QGP},\mu
;\varphi ,\psi ),
\end{equation}%
$M(\varphi ,\psi )$ is the weight function (Haar measure) given by: 
\begin{equation}
M(\varphi ,\psi )=\left( \sin \left( \tfrac{1}{2}(\psi +\tfrac{\varphi }{2}%
)\right) \sin (\tfrac{\varphi }{2})\sin \left( \tfrac{1}{2}(\psi -\tfrac{%
\varphi }{2})\right) \right) ^{2},
\end{equation}%
and $\widetilde{Z}$\ the generating function defined by:%
\begin{equation}
\widetilde{Z}(T,V_{QGP},\mu ;\varphi ,\psi )=Tr\left[ \exp \left( -\beta
\left( \widehat{H}_{0}-\mu \left( \widehat{N}_{q}-\widehat{N}_{\overline{q}%
}\right) \right) +i\varphi \widehat{I}_{3}+i\psi \widehat{Y}_{8}\right) %
\right] ,
\end{equation}%
where $\beta =\dfrac{1}{T}$ (with the units chosen as: $k_{B}=\hbar =c=1$), $%
\widehat{H}_{0}$\ is the free quark-gluon Hamiltonian, $\widehat{N}_{q}$ $%
\left( \widehat{N}_{\overline{q}}\right) $\ denotes the (anti-) quark number
operator, and $\widehat{I}_{3}$\ and $\widehat{Y}_{8}$\ are the color
\textquotedblleft isospin\textquotedblright\ and \textquotedblleft
hypercharge\textquotedblright\ operators respectively. Its final expression,
in the massless limit, can be put on the form:%
\begin{equation}
Z_{QGP}(T,V_{QGP},\mu )=\frac{4}{9\pi ^{2}}\int_{-\pi }^{+\pi }\int_{-\pi
}^{+\pi }d\varphi d\psi M(\varphi ,\psi )e^{V_{QGP}T^{3}g\left( \varphi
,\psi ,\frac{\mu }{T}\right) },
\end{equation}%
with: 
\begin{align}
g(\varphi ,\psi ,\frac{\mu }{T})& =\frac{\pi ^{2}}{12}(\frac{21}{30}d_{Q}+%
\frac{16}{15}d_{G})+\frac{\pi ^{2}}{12}\frac{d_{Q}}{2}\sum_{q=r,b,g}\left\{
-1+\left( \frac{\left( \alpha _{q}-i(\frac{\mu }{T})\right) ^{2}}{\pi ^{2}}%
-1\right) ^{2}\right\}  \notag \\
& \hspace*{-1.4cm}-\frac{\pi ^{2}}{12}\frac{d_{G}}{2}\sum_{g=1}^{4}\left( 
\frac{\left( \alpha _{g}-\pi \right) ^{2}}{\pi ^{2}}-1\right) ^{2},
\end{align}%
$d_{Q}=2N_{f}$\ and $d_{G}=2$\ being the degeneracy factors of quarks and
gluons respectively,$\ \alpha _{q}$ $\left( q=r,\,b,\,g\right) $\ the angles
determined by the eigenvalues of the color charge operators in eq. (9): 
\begin{equation}
\alpha _{r}=\tfrac{\varphi }{2}+\tfrac{\psi }{3},\;\alpha _{g}=-\tfrac{%
\varphi }{2}+\tfrac{\psi }{3},\;\alpha _{b}=-\tfrac{2\psi }{3},
\end{equation}%
and $\alpha _{g}$ $\left( g=1,...,4\right) $\ being: 
\begin{equation}
\alpha _{1}=\alpha _{r}-\alpha _{g},\;\alpha _{2}=\alpha _{g}-\alpha
_{b},\;\alpha _{3}=\alpha _{b}-\alpha _{r},\;\alpha _{4}=0.
\end{equation}

\section{Finite-size effects}

To study the effects of volume finiteness on the thermal QCD DPT within the
PCM, we'll examine in the following the behavior of some thermodynamic
quantities of the system with temperature, at a vanishing chemical potential 
$\left( \mu =0\right) $, considering the two lightest quarks $u$\ and $d$ $%
\left( N_{f}=2\right) $, and using the common value $B^{1/4}=145MeV$\ for
the bag constant.

The first quantity of interest for our study is the order parameter, which
can be simply in this case considered as represented by the hadronic volume
fraction. According to (3), its mean value is expressed as:%
\begin{equation}
<\mathfrak{h}(T,V)>=1-\frac{\int\limits_{-\pi }^{+\pi }\int\limits_{-\pi
}^{+\pi }d\varphi d\psi M(\varphi ,\psi )\int\limits_{0}^{1}\mathfrak{q}e^{%
\mathfrak{qR}\left( \varphi ,\psi ;T,V\right) }d\mathfrak{q}}{%
\int\limits_{-\pi }^{+\pi }\int\limits_{-\pi }^{+\pi }d\varphi d\psi
M(\varphi ,\psi )\int\limits_{0}^{1}e^{\mathfrak{qR}\left( \varphi ,\psi
;T,V\right) }d\mathfrak{q}},
\end{equation}%
with: $\mathfrak{R}\left( \varphi ,\psi ;T,V\right) =\left( g_{\mu
=0}(\varphi ,\psi )-\frac{\pi ^{2}}{30}-\frac{B}{T^{4}}\right) VT^{3}.$
After integration on the $\mathfrak{q}$ variable$,$\ the order parameter can
be written as:%
\begin{equation}
<\mathfrak{h}(T,V)>=\frac{L_{01}+L_{02}-L_{12}}{L_{01}-L_{11}},
\end{equation}%
where the general form of the integral terms appearing in this expression is:%
\begin{equation}
L_{nm}=\int\limits_{-\pi }^{+\pi }\int\limits_{-\pi }^{+\pi }d\varphi d\psi
M(\varphi ,\psi )\frac{\left( e^{\mathfrak{R}\left( \varphi ,\psi
;T,V\right) }\right) ^{n}}{\left( \mathfrak{R}\left( \varphi ,\psi
;T,V\right) \right) ^{m}}.
\end{equation}%
These integrals are then carried out using a suitable numerical method at
each fixed temperature and volume.

The second quantity of interest is the energy density $\varepsilon (T,V)$,
whose mean value is related to $<\mathfrak{h}(T,V)>$\ by:%
\begin{equation}
<\varepsilon (T,V)>=e_{HG}+\left( B-e_{HG}\right) \left( 1-<\mathfrak{h}%
(T,V)>\right) -3T^{4}\frac{\widetilde{L}_{11}-\widetilde{L}_{12}+\widetilde{L%
}_{02}}{L_{01}-L_{11}},
\end{equation}%
where the new integrals on $\varphi $\ and $\psi $\ , noted $\widetilde{L}%
_{nm},$\ are given by:%
\begin{equation}
\widetilde{L}_{nm}=\int_{-\pi }^{+\pi }\int_{-\pi }^{+\pi }d\varphi d\psi
M(\varphi ,\psi )g_{\mu =0}(\varphi ,\psi )\frac{\left( e^{\mathfrak{R}%
\left( \varphi ,\psi ;T,V\right) }\right) ^{n}}{\left( \mathfrak{R}\left(
\varphi ,\psi ;T,V\right) \right) ^{m}},
\end{equation}%
and: $e_{HG}=\frac{\pi ^{2}}{10}T^{4}.$

We can also examine the effects of volume finiteness by illustrating two
more quantities representing the first derivatives of the two previous ones,
i. e., the susceptibility $\chi $ defined as: 
\begin{equation}
\chi \left( T,V\right) =\dfrac{\partial \langle \mathfrak{h}\left(
T,V\right) \rangle }{\partial T},
\end{equation}
and the specific heat density $c\left( T,V\right) $ defined as:%
\begin{equation}
c\left( T,V\right) =\dfrac{\partial \langle \varepsilon \left( T,V\right)
\rangle }{\partial T}.
\end{equation}

Our results for the variations of these four response functions with
temperature at different system sizes are presented in the plots on Figs.
(1) and (2) and show a pronounced size dependence over almost the entire
temperature range. The curves show that in the limit of an infinite volume,
both $\langle \mathfrak{h}\rangle $ and $\dfrac{<\varepsilon >}{T^{4}}$
exhibit a finite sharp discontinuity, which is related to the latent heat of
the DPT, at a bulk transition temperature $T_{c}\left( \infty \right) \simeq
104.35MeV$, reflecting the first order character of the transition. The
quantity $\dfrac{\langle \varepsilon \rangle }{T^{4}}$ is traditionally
interpreted as a measure of the number of effective degrees of freedom; the
temperature increase causes then a \textquotedblleft
melting\textquotedblright\ of the constituent degrees of freedom
\textquotedblleft frozen\textquotedblright\ in the hadronic state, making
the energy density attain its plasma value. This finite discontinuity can be
mathematically described by a step function, which transforms to a $\delta $%
-function in $\chi $ and $c$. When the volume decreases, the four quantities
vary continuously such that the finite sharp jump is rounded off and the $%
\delta $-peaks are smeared out into finite peaks over a range of temperature 
$\delta T\left( V\right) $. This is due to the finite probability of
presence of the QGP phase below $T_{c}$ and of the hadron phase above $%
T_{c}, $ induced by the considerable thermodynamical fluctuations.

Another feature which can be noted is that the maxima of these rounded peaks
occur at effective transition temperatures $T_{c}\left( V\right) $ shifted
away from the bulk transition temperature for infinite volume $T_{c}\left(
\infty \right) $. This shift is a consequence of the color-singletness
requirement since the effective number of internal degrees of freedom for a
color-singlet QGP is drastically reduced\ with decreasing volume, as it can
clearly be seen from the curves and as it has been shown in \cite{EGR}.
Thus, the pressure of the QGP phase is lower at a given temperature, and the
mechanical Gibbs equilibrium between the two phases would then be reached
for $T_{c}(V)>T_{c}\left( \infty \right) .$

It can also be noted that, for decreasing volume, while the height of the
peak decreases, its width gets larger. To see this in more details, we
illustrate in the following the second derivative of the order parameter $%
\langle \mathfrak{h}\left( T,V\right) \rangle ^{\prime \prime }=\frac{%
\partial ^{2}\langle \mathfrak{h}\left( T,V\right) \rangle }{\partial T^{2}}%
, $ which reaches its extrema at temperatures $T_{1}(V)$ and $T_{2}(V)$. The
width of the transition region can simply be defined by the gap between
these two temperatures, i. e., $\delta T(V)=T_{2}(V)-T_{1}(V).$ The
variations of $\langle \mathfrak{h}\left( T,V\right) \rangle ^{\prime \prime
}$ with temperature for various sizes are illustrated in Fig. (3), from
which it can clearly be seen that the gap between the two extrema decreases
with increasing volume.

Finally from the obtained results for the finite size response functions,
four finite size effects can be observed:

\begin{itemize}
\item the rounding effect of the discontinuities

\item the smearing effect of the singularities

\item the shifting effect of the transition point

\item the broadening effect of the transition region,
\end{itemize}

and to study quantitatively the volume dependence of these effects, a
numerical scaling analysis is carried out, which will be presented in the
next section.

\section{Finite-size scaling analysis}

\subsection{Finite-Size Scaling}

In statistical mechanics, it is known that only in the thermodynamic limit
are phase transitions characterized by the appearance of singularities in
some second derivatives of the thermodynamic potential, such as the
susceptibility and the specific heat. For a first order phase transition,
the divergences are $\delta $-function singularities, corresponding to the
finite discontinuities in the first derivatives of the thermodynamic
potential, while for a second order phase transition the singularity has a
power-law form. For the thermal QCD DPT studied in this work, $\delta $%
-singularities appear in the susceptibility $\chi $ and in the specific heat
density $c$ at the thermodynamic limit. In finite volumes, these $\delta $%
-functions are found to be smeared out into finite peaks. To the finite size
effects, four useful characteristic quantities can be associated as
illustrated on Fig. (4), which are the maxima of the peaks of the
susceptibility $\chi _{T}^{\max }\left( V\right) $ and the specific heat
density $c_{T}^{\max }\left( V\right) $, the shift of the transition
temperature $\tau _{T}(V)=T_{c}\left( V\right) -T_{c}\left( \infty \right) $
and the width of the transition region $\delta T\left( V\right) .$ Each of
these quantities can be considered as a signature which may anticipate the
behavior in the thermodynamic limit, and is expected to present a scaling
behavior described by a power law of the volume characterized by a \textit{%
Scaling Critical Exponent. }For a first order phase transition, the set of
power laws is: 
\begin{equation}
\left\{ 
\begin{array}{c}
\chi _{T}^{\max }\left( V\right) \sim V^{\gamma
}\;\;\;\;\;\;\;\;\;\;\;\;\;\;\;\;\ \;\;\;\;\;\;\;\ \ \ \  \\ 
c_{T}^{\max }\left( V\right) \sim V^{\alpha
}\;\;\;\;\;\;\;\;\;\;\;\;\;\;\;\;\;\ \;\;\;\;\;\;\;\ \ \ \ \  \\ 
\delta T\left( V\right) \sim V^{-\theta
}\;\;\;\;\;\;\;\;\;\;\;\;\;\;\;\;\;\;\;\ \;\;\;\;\;\;\;\;\;\;\  \\ 
\tau _{T}(V)=T_{c}\left( V\right) -T_{c}\left( \infty \right) \sim
V^{-\lambda }\ ,%
\end{array}%
\right.
\end{equation}%
and it has been shown in the FSS theory \cite{FBe82, CLB86, BH88} that in
this case, the scaling exponents $\theta $, $\lambda $, $\alpha $ and $%
\gamma $\ are all equal to unity, and it is only the dimensionality which
controls the finite size effects.

At a second order phase transition, the correlation length diverges as: $\xi
\propto \left\vert T-T_{c}\right\vert ^{-\nu }$, and then we predict the
same power law behavior as for a first order transition, but with different
scaling critical exponents, usually given as:%
\begin{equation}
\left\{ 
\begin{array}{c}
\chi _{T}^{\max }\left( V\right) \sim V^{\gamma /\nu
}\;\;\;\;\;\;\;\;\;\;\;\;\;\;\;\;\ \;\;\;\;\;\;\;\ \ \ \  \\ 
c_{T}^{\max }\left( V\right) \sim V^{\alpha /\nu
}\;\;\;\;\;\;\;\;\;\;\;\;\;\;\;\;\;\ \;\;\;\;\;\;\;\ \ \ \ \  \\ 
\delta T\left( V\right) \sim V^{-1/\nu
}\;\;\;\;\;\;\;\;\;\;\;\;\;\;\;\;\;\;\;\ \;\;\;\;\;\;\;\;\;\;\  \\ 
\tau _{T}(V)=T_{c}\left( V\right) -T_{c}\left( \infty \right) \sim V^{-1/\nu
}\ .%
\end{array}%
\right.
\end{equation}

The values of the scaling critical exponents may then give an indication on
the order of a PT and are usually used as a criterion for the determination
of this latter \cite{Or96,HEN99}.

\subsection{Numerical Determination of the Scaling Critical Exponents for
the Thermal DPT}

In the following, we use a FSS analysis to recover the scaling exponents $%
\theta $, $\lambda $, $\alpha $ and $\gamma $ for the thermally driven DPT.
For this purpose, we proceed by studying the behavior of the
response-function maxima, their rounding as well as the shift of the
effective transition temperature with varying volume. Let's note that the
determination of the location of the maxima of the finite size peaks as well
as their heights is done in a numerical way, and this yields a systematic
error, which is estimated and given for the determined scaling exponents.

\subsubsection{\textit{Susceptibility, Specific Heat and Smearing Scaling
Exponents}}

The data of the maxima of the rounded peaks of the susceptibility $\mid \chi
_{T}\mid ^{\max }\left( V\right) $\ and the specific heat density $%
c_{T}^{\max }\left( V\right) $\ are plotted versus volume in Figs. (5-left)
and (5-right) respectively, and their linearity with $V$ can clearly be
noted. A numerical parametrization with the power-law forms: $\mid \chi
_{T}\mid ^{\max }\left( V\right) \sim V^{\gamma }$ and: $c_{T}^{\max }\left(
V\right) \sim V^{\alpha },$ gives the values\ of the susceptibility scaling
exponent: $\gamma =1.01\pm 0.03,$ and the specific heat scaling exponent: $%
\alpha =1.007\pm 0.031,$ where the associated errors are systematic ones.

Fig. (6) illustrates the plot of the results of the width $\delta T\left(
V\right) $ with the inverse of the volume, and their fit to the power law
form: $\delta T\left( V\right) \sim V^{-\theta }.$ The obtained smearing
scaling exponent is: $\theta =1.03\pm 0.03.$

\subsubsection{\textit{The Shift Scaling Exponent}}

For the study of the shift of the transition temperature $\tau
_{T}(V)=T_{c}\left( V\right) -T_{c}\left( \infty \right) $, we need to
locate the effective transition temperature\ in a finite volume $T_{c}\left(
V\right) $. A way to define $T_{c}\left( V\right) $ is as being the location
of the maxima of the rounded peaks of the susceptibility and the specific
heat, shifted away from the true transition temperature $T_{c}\left( \infty
\right) $. Results of the shift of the transition temperature obtained in
this way are plotted in Fig.(7) versus inverse volume. The shift critical
exponent obtained from a fit to the form: $\tau _{T}(V)\sim V^{-\lambda }$,
is: $\lambda =0.876\pm 0.041.$

\subsubsection{\textit{The Shift from the Fourth Binder Cumulant}}

Another way for locating $T_{c}\left( V\right) $\ is to consider the fourth
order cumulant of the order parameter proposed in \cite{CLB86} and defined
as:%
\begin{equation}
B_{4}\left( T,V\right) =1-\dfrac{\langle \mathfrak{h}^{4}\left( T,V\right)
\rangle }{3\ \langle \mathfrak{h}^{2}\left( T,V\right) \rangle ^{2}},
\end{equation}%
which presents a minimum value at an effective transition temperature whose
shift from the true transition temperature is of order $V^{-1}$ for a first
order transition. This cumulant has been proven to be a suitable indicator
of the order of the transition in a finite volume, since\ a nonvanishing
value of $\left( \frac{2}{3}-B_{4}^{\ \min }\right) $\ signals a first order
transition, whereas for a second order transition, $\underset{V\rightarrow
\infty }{lim}\left( \frac{2}{3}-B_{4}\right) =0$ even at the transition
point. Indeed, for a first order transition the quantity $\underset{%
V\rightarrow \infty }{lim}\left( \frac{2}{3}-B_{4}\right) $ vanishes at all
points apart from the transition point, and $\left( \frac{2}{3}-B_{4}^{\
\min }\right) $\ measures the latent heat \cite{CLB86, BH88,Bind97,LB2000}.

The expression of $B_{4}\left( T,V\right) $ as function of the integral
terms, for this case of the deconfinement transition,\ is\ after calculation:%
\begin{equation}
B_{4}(T,V)=1-\frac{\left( L_{11}-L_{01}\right) \left( 24\left(
L_{15}-L_{05}-L_{04}\right) -12L_{03}-4L_{02}-L_{01}\right) }{3\ \left(
2L_{13}-2L_{03}-2L_{02}-L_{01}\right) ^{2}}.
\end{equation}

Fig.(8-left) illustrates the variations of the fourth cumulant of the order
parameter with temperature for various volumes, and shows that the locations
of the minima in finite sizes $T_{\min }\left( V\right) $ are shifted to
higher values from $T_{c}\left( \infty \right) $. Data of the shift of the
transition temperature obtained in this way are plotted in Fig.(8-right)
versus inverse volume, and the scaling shift critical exponent obtained from
a fit to the form: $\tau _{T}^{\prime }(V)=T_{\min }\left( V\right)
-T_{c}\left( \infty \right) \sim V^{-\lambda ^{\prime }}$, is: $\lambda
^{\prime }=0.883\pm 0.043.$

\subsubsection{Parametrization of the Order Parameter and the Susceptibility}

Within the model used in this work, the order parameter in the limit of
infinite volume being equal to $1$ below the transition temperature and zero
above, it can then be expressed in a simple way using the Heaviside
step-function as: 
\begin{equation}
\langle \mathfrak{h}\left( T,V\longrightarrow \infty \right) \rangle
=1-\Theta \left( T-T_{c}\left( \infty \right) \right) .
\end{equation}

Such a parametrization has been used in \cite{BO87}, for a similar case of a
system of coexisting hadronic matter and QGP phases. Using one of the known
mathematical representations of the smoothed step function $\Theta \left(
T-T_{c}\right) ,$\ the order parameter may then be expressed as: 
\begin{equation}
\langle \mathfrak{h}\left( T,V\right) \rangle =\frac{1}{2}\left( 1-\tanh
\left( \frac{T-T_{c}\left( V\right) }{\Gamma _{T}\left( V\right) }\right)
\right) ,
\end{equation}%
where $T_{c}$\ is the effective transition temperature and $\Gamma _{T}$\
the half-width of the rounded transition region, which leads to the
susceptibility expression :%
\begin{equation}
\chi \left( T,V\right) =\frac{-1}{2\Gamma _{T}\left( V\right) ~\cosh
^{2}\left( \frac{T-T_{c}\left( V\right) }{\Gamma _{T}\left( V\right) }%
\right) }.
\end{equation}

The parametrisation choice (26) is the most accepted physically, and it can
be understood phenomenologically in the context of the Double Gaussian Peaks
model where the obtained expressions of the order parameter and the
susceptibility are very similar to (26) and (27) \cite{Kbinder,Or96}.

An illustration of such parametrizations of the order parameter at the
volume $V=4000fm^{3}$, and the susceptibility at the volume $V=900fm^{3}$
are presented in Figs.(9-left) and (9-right) respectively, and the
parameters $T_{c}\left( V\right) $\ and $\Gamma _{T}\left( V\right) $\
obtained from each fit are given.

The results for the width of the transition region $\delta T^{~fit(1)}\left(
V\right) $\ and the shift of the transition temperature $\tau
_{T}^{~fit(1)}\left( V\right) $\ from the parametrization of the order
parameter are plotted with inverse volume in Fig.(10) and those from the
parametrization of the susceptibility, $\delta T^{~fit(2)}\left( V\right) $\
and $\tau _{T}^{~fit(2)}\left( V\right) $ in Fig.(11). Their fits to power
law forms: $\tau _{T}^{~fit(1)}(V)\sim V^{-\lambda _{1}},$ $\tau
_{T}^{~fit(2)}(V)\sim V^{-\lambda 2},$ $\delta T^{~fit(1)}\left( V\right)
\sim V^{-\theta _{1}}$ and $\delta T^{~fit(2)}\left( V\right) \sim
V^{-\theta _{2}}$\ give: $\lambda _{1}=0.830\pm 0.013~,$ $\lambda
_{2}=0.857\pm 0.006~,$ $\theta _{1}=0.990\pm 0.015\ $and $\theta
_{2}=1.032\pm 0.003\ .$

\section{Conclusion}

Our work has shown the influence of the finiteness of the system size on the
behavior of some response functions in the vicinity of the transition point.
The sharp transition observed in the thermodynamical limit, signaled by
discontinuities in the order parameter and in the energy density at a
transition temperature $T_{c}\left( \infty \right) $, is rounded off in
finite volumes, and the variations of these thermodynamic quantities are
perfectly smooth on the hole range of temperature. The delta function
singularities appearing in the first derivatives of these discontinuous
quantities, i.e., in the susceptibility and specific heat density, are then
smeared out into finite peaks of widths $\delta T\left( V\right) .$ The
maxima of these peaks occur at effective transition temperatures $%
T_{c}\left( V\right) $\ shifted away from the true transition temperature $%
T_{c}\left( \infty \right) $. A FSS analysis of the behavior of the maxima
of the rounded peaks of the susceptibility $\chi _{T}^{\max }\left( V\right) 
$\ and the specific heat density\ $c_{T}^{\max }\left( V\right) $, the width
of the transition region$\ \delta T\left( V\right) $, and the shift of the
effective transition temperature relative to the true one $\tau _{T}\left(
V\right) =T_{c}\left( V\right) -T_{c}\left( \infty \right) $, shows their
power-law variations with the volume characterized by the scaling critical
exponents $\gamma ,\,\alpha ,\,\theta ,$ and $\lambda $ respectively.
Numerical results for these scaling exponents are obtained and are in good
agreement with our analytical results: $\gamma =\alpha =\theta =\lambda =1$
obtained in \cite{LYA02}, except for the shift critical exponent which
slightly deflects from the analytical value $1$. This may be due to the
difficulty of locating accurately the peaks of $\chi $, $c$\ and $B_{4},$
especially in the case of large volumes for which the peaks become very
sharp. However, a first estimate of the scaling critical exponents for the
thermal DPT within this model has been obtained, and our result expressing
the equality between all scaling exponents and the space dimensionality is a
consequence of the singularity characterizing a first order phase transition
and agrees very well with the predictions of other FSS theoretical
approaches \cite{BDT2000} and with the results of both\ lattice QCD \cite%
{BKLP99} and Monte Carlo models calculations.

\newpage

{\LARGE Figure Captions\vspace*{1cm}}

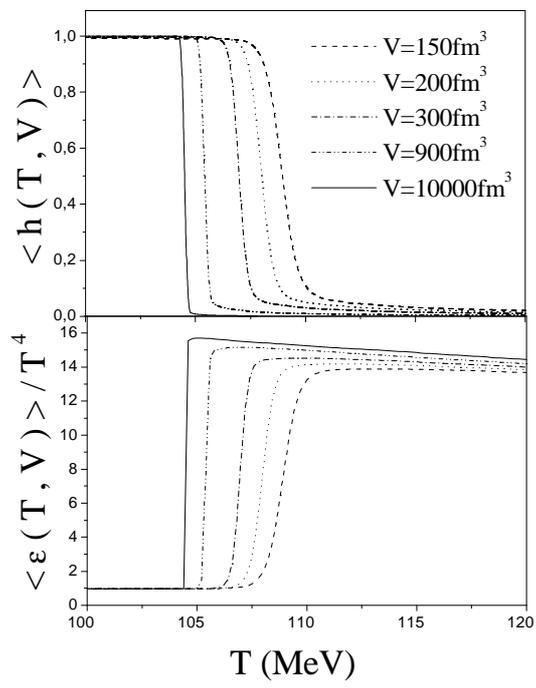
Fig. 1: Temperature variation of the order parameter (top) and the energy
density normalized by $T^{4}$\ (bottom) at $\mu =0,$ for different system
volumes.

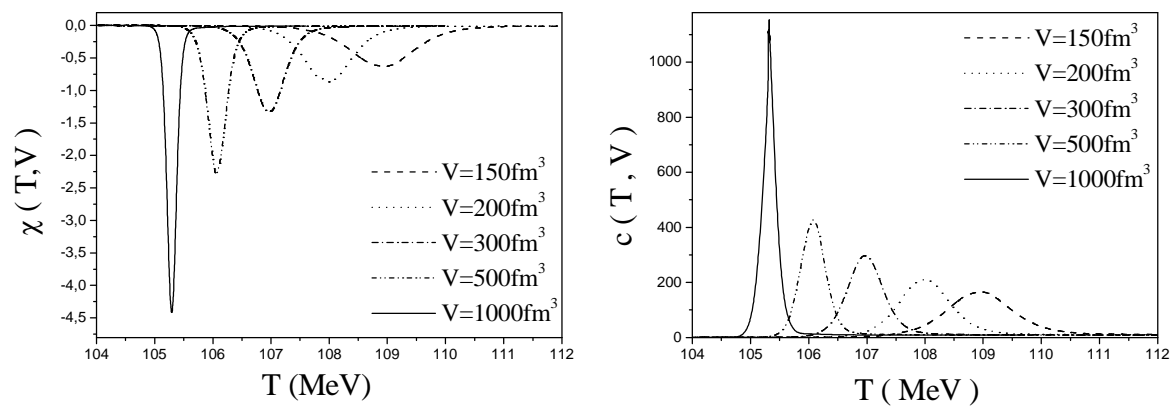
Fig. 2: Plot of (Left) the susceptibility $\chi \left( T,V\right) $ and
(Right) the specific heat density $c\left( T,V\right) ,$ versus temperature
for different system volumes.

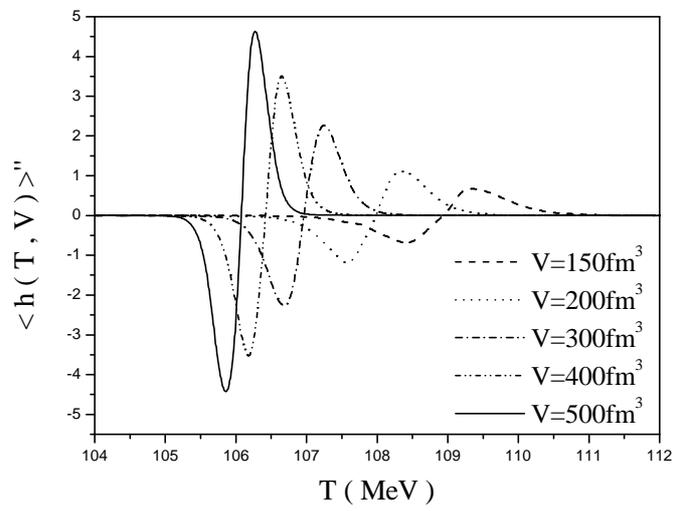
Fig. 3: Plot of the second derivative of the order parameter $\langle 
\mathfrak{h}\left( T,V\right) \rangle ^{\prime \prime }=\frac{\partial
^{2}\langle \mathfrak{h}\left( T,V\right) \rangle }{\partial T^{2}}$ versus
temperature for different system volumes.

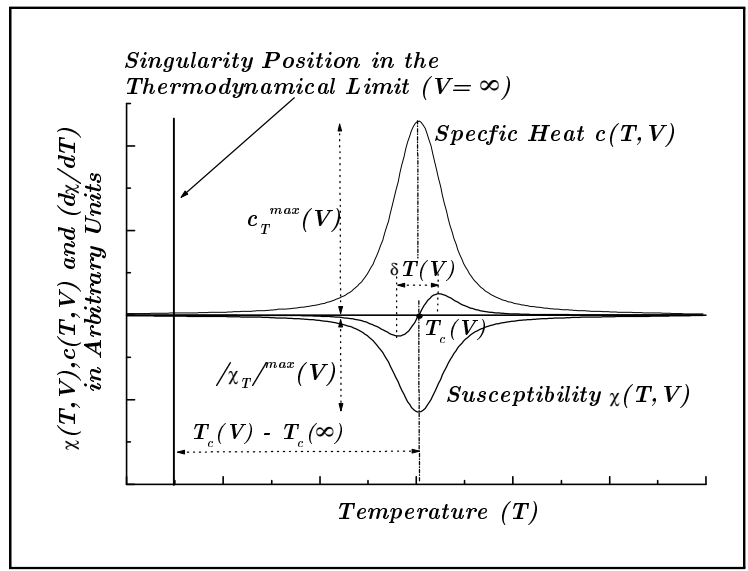
Fig. 4: Illustration of the finite size behavior of the susceptibility $\chi
\left( T,V\right) $, the specific heat density $c\left( T,V\right) $\ and
the second derivative of the order parameter $\partial \chi /\partial T$.

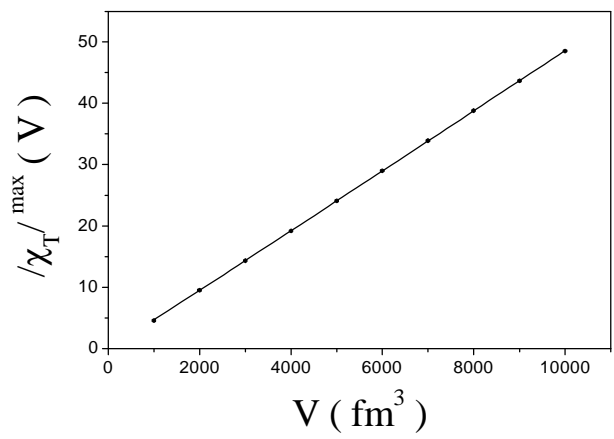
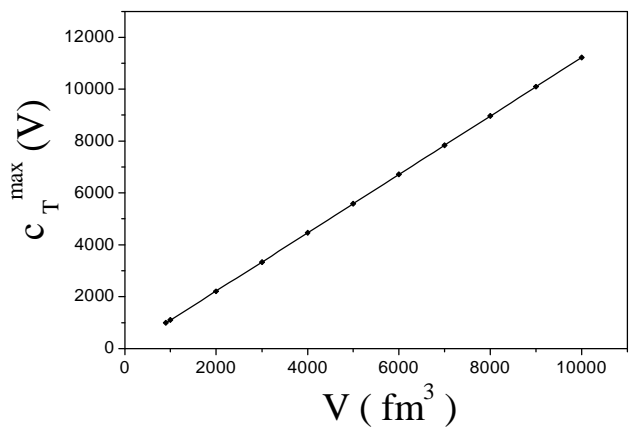
Fig. 5: Variation of (Left) the susceptibility maxima $\mid \chi _{T}\mid
^{\max }\left( V\right) $\ and (Right) the specific heat density maxima $%
c_{T}{}^{\max }\left( V\right) $ with volume.

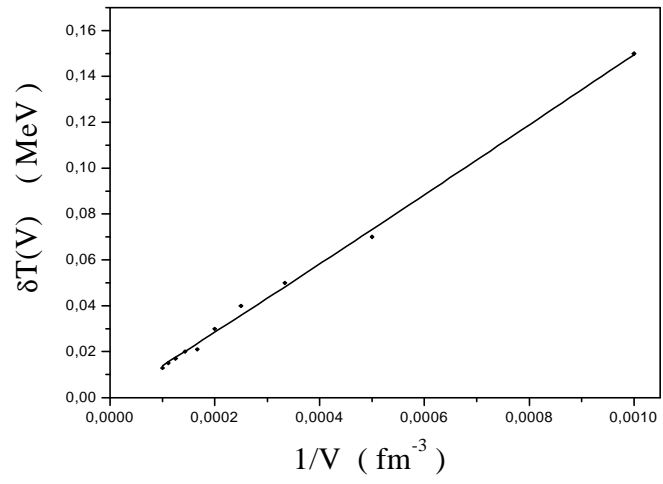
Fig. 6: Variation of the width of the temperature region over which the
transition is smeared $\delta T(V)$ with inverse volume.

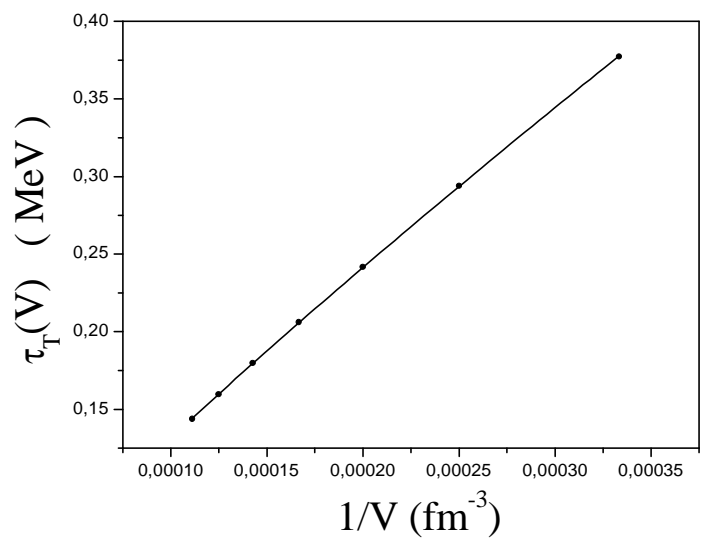
Fig. 7: Variation of the shift of the transition temperature $\tau _{T}(V)$
(from the maxima of $\chi \left( T,V\right) \ $and $c\left( T,V\right) $)
with inverse volume.

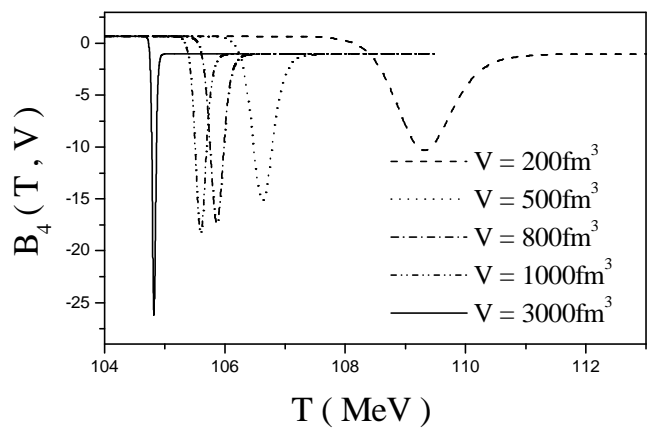
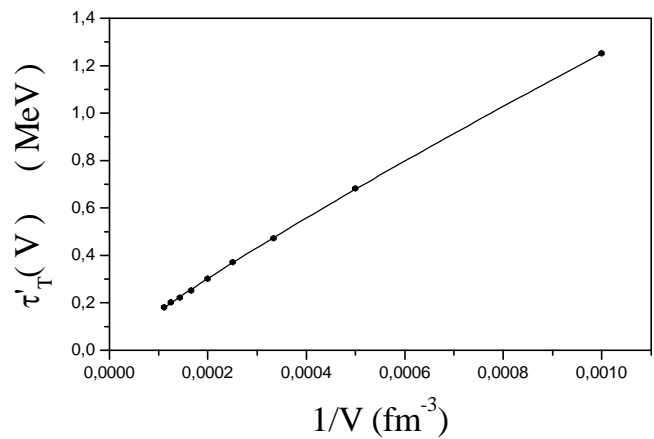
Fig. 8: (Left) Plot of the fourth Binder cumulant $B_{4}\left( T,V\right) $
versus temperature for different system volumes, and (Right) variation of
the shift of the transition temperature $\tau _{T}^{\prime }(V)$ (from the
minimum of $B_{4}\left( T,V\right) $) with inverse volume.

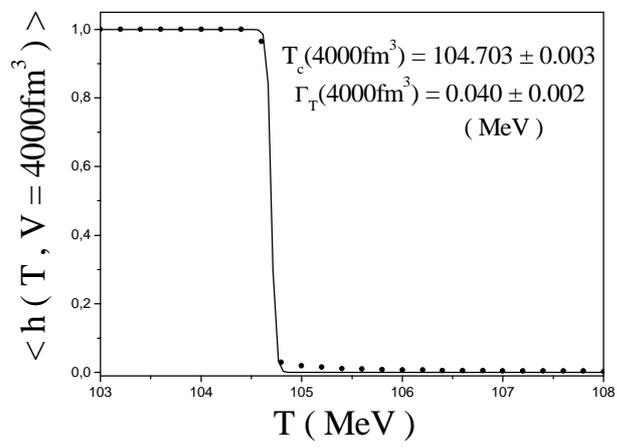
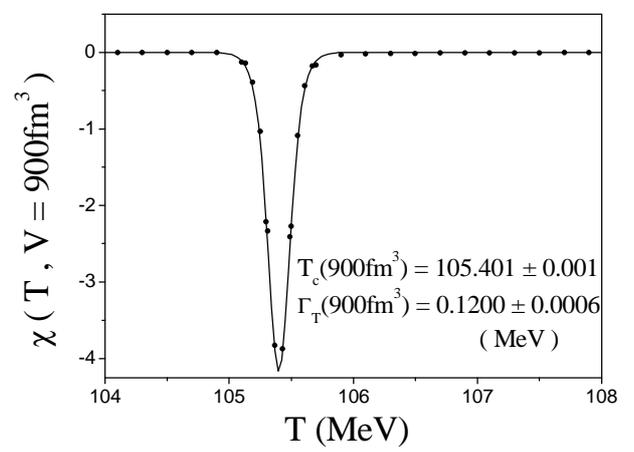
Fig. 9: Illustration of the fits of the order parameter to the form (26) at
the volume $V=4000fm^{3},$ and of the susceptibility to the form (27) at the
volume $V=900fm^{3}.$

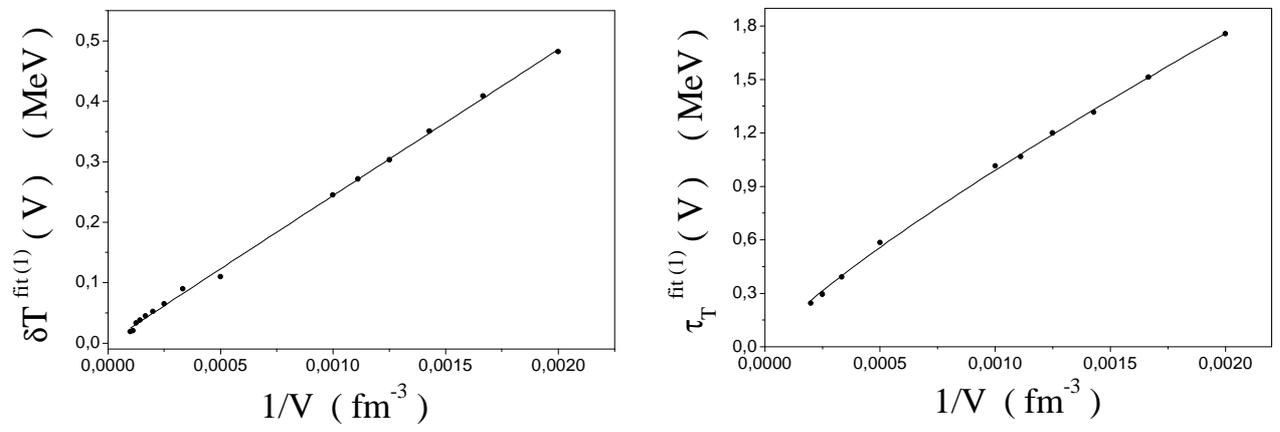
Fig. 10: Variations of (Left) the width of the transition region $\delta
T^{fit(1)}(V)$, and (Right) the shift of the transition temperature $\tau
_{T}^{fit(1)}(V)$, obtained from the parametrization in eq. (26), with
inverse volume.

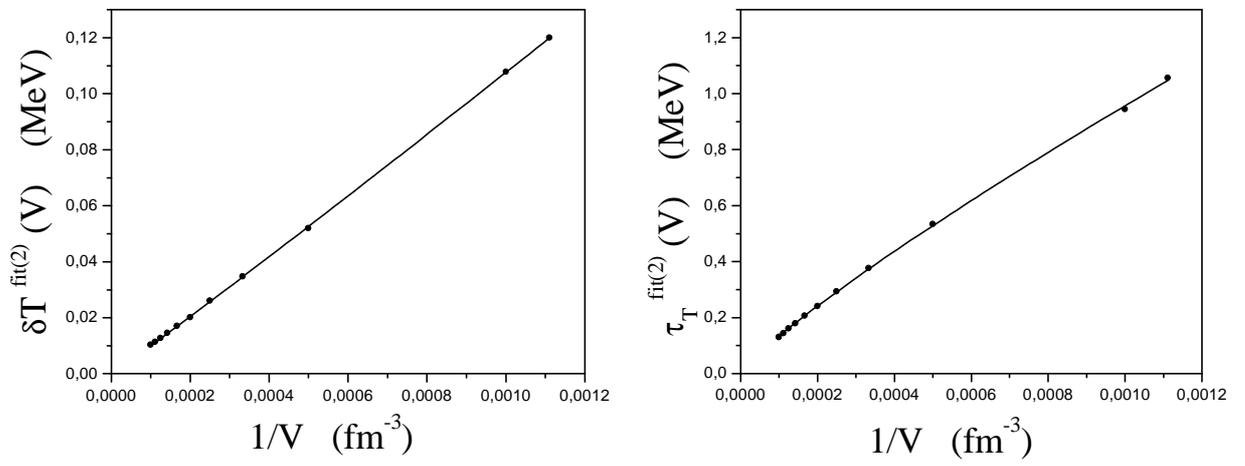
Fig. 11: Variations of (Left) the width of the transition region $\delta
T^{fit(2)}(V)$, and (Right) the shift of the transition temperature $\tau
_{T}^{fit(2)}(V)$, obtained from the parametrization in eq. (27), with
inverse volume.


\begin{thebibliography}{99}
\bibitem{LeeYang52} C. N. Yang and T. D. Lee, Phys. Rev. \textbf{87} (1952)
404; Phys. Rev. \textbf{87} (1952) 410.

\bibitem{Im80} J. Imry, Phys. Rev. B \textbf{21} (1980) 2042.

\bibitem{BL84} K. Binder and D. P. Landau, Phys. Rev. B \textbf{30} (1984)
1477.

\bibitem{SSG98} C. Spieles, H. St\"{o}cker and C. Greiner, Phys. Rev. C 
\textbf{57} (1998) 908.

\bibitem{Or96} H. Meyer-Ortmanns, Rev. Mod. Phys. \textbf{68} (1996) 473.

\bibitem{RT81} K. Redlich and L. Turko, Z. Phys. C \textbf{5} (1980) 201;
L.Turko, Phys. Lett.\ \textbf{104B} (1981) 153.

\bibitem{EGR} H.-Th. Elze, W. Greiner and J.\ Rafelski , Phys. Lett.\ 
\textbf{124B} (1983) 515; Zeit. Phys. C \textbf{24} (1984) 361; H.-Th. Elze
and W. Greiner, Phys. Lett. \textbf{179B} (1986) 385.

\bibitem{TLR} A. Tounsi, J. Letessier and J. Rafelski, hep-ph/9811290.

\bibitem{MSS} M. G. Mustafa, D. K. Srivastava and B. Sinha, Eur. Phys. J. C 
\textbf{5} (1998) 711; nucl-th/9712014.

\bibitem{YG02} G. Yezza, Magister thesis in theoretical physics, Ecole
Normale Sup\'{e}rieure-Kouba, Algiers (March 2002).

\bibitem{FBe82} M. E. Fisher and A. N. Berker, Phys. Rev. B \textbf{26}
(1982) 2507.

\bibitem{CLB86} M. S. Challa, D. P. Landau and K. Binder, Phys. Rev. B 
\textbf{34} (1986) 1841.

\bibitem{BH88} K. Binder and D. W. Heermann, \textit{Monte Carlo Simulations
in Statistical Physics}, (Springer-Verlag, 1988, 2nd ed. 2002).

\bibitem{HEN99} M. Henkel, \textit{Conformal Invariance and Critical
Phenomena}, Springer Verlag (1999).

\bibitem{Bind97} K. Binder, Rep. Prog. Phys. \textbf{60} (1997) 487.

\bibitem{LB2000} D. P. Landau and K. Binder, \textit{A Guide to Monte Carlo
Simulation in Statistical Physics, }(Cambridge University Press, 2000).

\bibitem{BO87} J. P. Blaizot and J. Y. Ollitraut, Phys. Rev. \textbf{D36}
(1987) 916.

\bibitem{Kbinder} K. Binder,\ Phys. Rev. Lett. \textbf{47} (1981) 693; Zeit.
Phys. B \textbf{43} (1981) 119; K. Binder \& D. P. Landau, Phys. Rev. B 
\textbf{30} (1984) 1477.

\bibitem{LYA02} M. Ladrem, A. Ait-El-Djoudi and G. Yezza, communication at
the international conference `Quark Confinement and the Hadron spectrum',
held in Gargnano-Italy from 10 to 14 September 2002.

\bibitem{BDT2000} J. G. Brankov, D. M. Danchev and N. S. Tonchev,\textit{\
Theory of Critical Phenomena in Finite-Size Systems - Scaling and Quantum
Effects}, (World Scientific, 2000).

\bibitem{BKLP99} B. Beinlich, F. Karsch, E. Laermann and A. Peikert, Eur.
Phys. J. C \textbf{6 }(1999) 133.\newpage
\end{thebibliography}
\end{document}